\documentclass[preprint,showpacs]{revtex4}

\usepackage{amsfonts,amssymb,amsmath,amsthm,amscd,epsf}
\usepackage{amscd}
\usepackage[dvips]{graphicx,psfrag}
\usepackage{epsf,psfig}

\newcommand{\BT}{{\mathbf{T}}}
\newcommand{\bS}{\mathbf{S}}

\theoremstyle{plain}

\theoremstyle{definition}

\theoremstyle{remark}

\begin{document}

\title{How to compute using globally coupled oscillators}

\author{Peter Ashwin and Jon Borresen}
\affiliation{Department of Mathematical Sciences,
 University of Exeter,
 Exeter EX4 4QE, UK}

\date{\today}

\begin{abstract}
Synchronization is known to play a vital role within many highly
connected neural systems such as the olfactory systems of fish and insects.
In this paper we show how one can robustly and effectively
perform practical computations using small perturbations to a very
simple globally coupled network of coupled oscillators. Computations
are performed by exploiting the spatio-temporal dynamics of a
robust attracting heteroclinic network (also referred to as
`winnerless competition' dynamics). We use different cluster synchronization states
to encode memory states and use this to design a simple multi-base counter.
The simulations indicate that this gives a robust computational
system exploiting the natural dynamics of the system.
\end{abstract}



\maketitle

Recent experimental and theoretical work by several authors has highlighted the
crucial role of spatio-temporal dynamics in neural systems. In particular,
work on olfactory systems of insects and fish has demonstrated that recognized odors
manifest themselves as different spatio-temporal outputs from certain sub-systems such
as \cite{Rab&al01,Mar&Hug03,Bur&al02,Huerta} for insect antennal lobes (AL) and
\cite{Fri&al04a,Fri&al04b} for the Zebrafish olfactory bulb (OB).
It has been suggested that the dynamics of the network responsible for the
transformation of input to output of the AL/OB for these systems is one of `winnerless competition'
\cite{Lau&al01,Afr04,Sel&al03}; this is a class of attracting dynamics consisting of
saddle periodic orbits linked together by unstable manifolds.
Such attractors may seem at first unnatural but in systems that are
close to symmetric and in systems that have invariant subspaces they may be robust;
i.e. they may persist under arbitrary perturbations that preserve the symmetric (or
invariant subspace) structure. Such attractors have been found in a range
of physically relevant models and experiments \cite{Kru97,AshHets}.

The AL/OB system is far from being a simple relay; it shows
effects such as short term memory, anticipation and computation \cite{Lau&al01}.
The aim of this paper is to show that globally coupled systems idealizing the AL/OB
architecture can be used to explicitly design computational systems by 
exploiting robust heteroclinic attractors (winnerless competition).
We use globally coupled phase oscillators in regimes where one finds
slow switching between cluster states \cite{Han&al93a,Han&al93b,Kor&al01,Kor03}.
Previous work has shown \cite{Ash&Bor04} that robust heteroclinic attractors for
a particular system of five globally coupled phase oscillators allow one to encode
up to twenty different memory states as synchronized clusters, and moreover to
effectively move between them by means of small perturbations. The number of such
states scales exponentially with $N$ the number of oscillators. In this paper
we go further to show that one can in principle perform arbitrary computations
using a globally coupled cluster for `memory/timing' on addition 
of perturbations that may have very low amplitude.

We consider perturbations to the 
model of Hansel {\it{et al.}} \cite{Han&al93a} for $N$ globally coupled
oscillators with phases $\theta_i \in \BT=[0,2\pi)$ given by
\begin{equation} \label{eqhansel}
\dot{\theta_i} = \omega+ \frac{1}{N}
\sum^N_{j=1} g\left(\theta_i-\theta_j\right)+\eta w_i(t)+ \epsilon I_i(t),
\end{equation}
when $i=1,\cdots,N$ and $g(\phi) = -\sin (\phi+\alpha)+r \sin (2\phi)$; see
also \cite{Han&al93b,Kor&al01,Kor03}. The quantities $w_i(t)$ represent
derivatives of independent, identically distributed Brownian processes
with zero mean and unit variance per unit time. The inputs $|I_i(t)|\leq 1$
are used to control the state of the system. We include $\eta$ and $\epsilon$,
small parameters that control the strength of the noise and perturbations
respectively.

The system (\ref{eqhansel}) for the unperturbed system ($\eta=\epsilon=0$) has
symmetry $\bS_N$ corresponding to permutations of the $N$ oscillators. Using these
symmetries one can identify robust heteroclinic attractors (first noted theoretically in
\cite{AshHets} and then practically for this example in \cite{Han&al93a})
that exist for open sets of parameters. For certain parameter values the attractors
consist of saddle periodic orbits of symmetry $\bS_k\times \bS_{N-k}$,
i.e. states where all oscillators are in one of two possible phases at any point in
time and such that precisely $k$ of them are in one phase. The system also has an
$\bS^1$ phase-shift symmetry that allows one to characterize periodic orbits as
group orbits.  For
\begin{equation} \label{eqn:params}
N=5, ~\alpha=1.25, ~\omega=5, ~r=0.25,
\end{equation}
the only attractor for
unperturbed (\ref{eqhansel}) consists of twenty periodic orbits with symmetry
$\bS_2\times \bS_3$ and their unstable invariant manifolds. This forms them into a single
strongly connected network.

\begin{figure}
\centerline{\mbox{\psfig{file=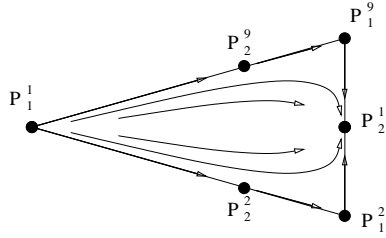,width=50mm}}}
\caption{\label{fig_man}
Schematic showing part of the two dimensional unstable manifold $W^u(P_1^1)$;
almost all trajectories leaving from $P_1^1$ converge to $P_2^1$ but exceptional
trajectories converge to $P_2^l$ for three possible $l\neq 1$, two of which are shown
here.
}
\end{figure}

The periodic
orbits involved are $\bS^1$ orbits of $x_j=(0,0,0,\psi_j,\psi_j)$ with $j=1,2$ and
permutations thereof. We list a set of permutations $\sigma_k$
with $k=1,\cdots,10$ that map the periodic orbit through $x_j$ onto the 10
possible symmetric images and denote by $P_j^k$ the periodic orbit given by the
$\bS^1$ orbit of $\sigma_k x_j$ for $j=1,2$ and $k=1,\cdots,10$. These periodic
orbits are listed in Table~\ref{tab_pos}.
For the given parameters (\ref{eqn:params}), $\psi_1=1.339$ and $\psi_2=0.799$ \cite{Ash&Bor04}. If
we write
$\Sigma= \bigcup_{i=1}^{10}\bigcup_{j=1}^2 W^u(P_j^k)$ where $W^u(P)$ is the unstable
manifold of the saddle periodic orbit $P$ then any randomly chosen initial condition approaches
$\Sigma$; this manifests itself as a sequence of `slowing down' switchings between different cluster
states $P_j^k$.
Figure~\ref{fig_man} shows a detail of the unstable manifold of
one of the $P_1^k$.

For larger $N$ (\ref{eqhansel}) displays heteroclinic cycles between
$\bS_k\times \bS_{N-k}$ symmetry states for a variety of $k$ with
$N/3<k\leq N/2$ \cite{Kor&al01};
this gives rise to more complicated attracting heteroclinic networks that are
currently under investigation by the authors. The unperturbed dynamics is such
that after a transient period the system state is near one of the $P_j^k$ except for
increasingly occasional short transitions. As $\Sigma$ is asymptotically stable this
is also true even for the perturbed system as long as the perturbation in small enough.
We exploit this feature for our computational system.
Typical initial conditions for the unperturbed system lead to an asymptotic state that
is a slowing-down oscillation between $P_1^k$ and $P_2^k$ for some $k$ determined
by the initial condition.

Trajectories of the perturbed system may visit many or all $P_j^k$ where $j$ always alternates
between $1$ and $2$ and $k$ changes
on a longer timescale \cite{Ash&Bor04}.
This is because the unstable manifold of $P_1^k$ is two dimensional and only exceptional
trajectories on this connect to $P_2^l$ with $k\neq l$. In the presence of
noise we refer to the average time of switching between $P_j^k$ as the {\em cycling
time} $T_c$; one can show that this scales as $-\ln \eta$ \cite{Han&al93a}. The average time of
switching between $P_j^k$ with different $k$ we refer to
as the {\em switching time} $T_s$.
The order parameter $\chi= \frac{1}{N} \left|\sum_{k=1}^N e^{i\theta_k}\right|$
often considered for such systems \cite{Han&al93a,Kor&al01} oscillates at a rate
given by the cycling time. Because this measure is invariant under permutations of the phases it is
not possible to detect switches and compute $T_s$
using only $\chi$.

\begin{figure}
\centerline{\mbox{\psfig{file=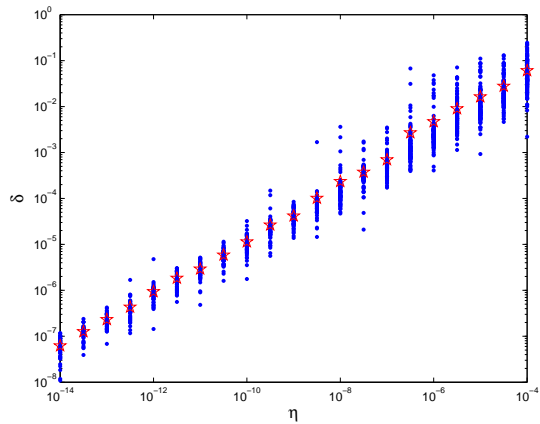,width=8cm}}}
\caption{\label{fig:epsballs1}
Dots show the value of $\delta=\inf\{Y^k_j(t)~:~k,j,t\in[t_0,t_0+T_c)\}$,
the smallest value of any $Y^k_j$ during a cycle time $T_c$ in the presence
of noise with amplitude $\eta$, after transients have decayed and for
a range of random initial conditions.  The mean closest approach (taking
100 cycles) scales as $\langle \delta \rangle=O(\sqrt{\eta})$.
}
\end{figure}

\begin{figure}
\centerline{\mbox{\psfig{file=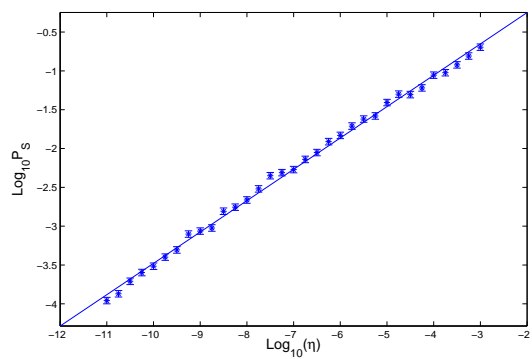,width=8cm}}}
\caption{\label{fig:noise}
Probability of switching $P_s$ at each cycle for specified $\eta$ with
average over 100 trials. Error bars show two standard deviations and a least squares fit line is
also included. This fits $P_s\sim \sqrt{\eta}$.
}
\end{figure}

One can view the globally coupled system (\ref{eqhansel}) as
a computational system with five inputs $I_i(t)$, $i=1..5$ and
twenty outputs $Y_j^k(\theta) = |\theta-P_j^k|_1$ (where
$|\theta|_1=\sum_i |\theta_i|$), namely if $Y_j^k$ is small then the
trajectory is close to $P_j^k$.
Figure \ref{fig:epsballs1} shows the statistics of $\delta$, the local
minimum of the $Y_j^k$ over one cycle. The mean $\langle \delta\rangle$
varies proportionally as
the square root of the noise, reminiscent of the algebraic scaling in
heteroclinic switching rates with noise level found in \cite{Armbruster}.
Defining the mean probability of switching per cycle $P_s$, we can estimate
$P_s= T_c/T_s$. Numerical simulations indicate that $T_s/T_c= O(\eta^{1/2})$
as $\eta\rightarrow 0$; see Figure~\ref{fig:noise}, where we detect
close approaches by $|\theta-P_j^k|_1<\Delta$ where $\Delta=10\sqrt{\epsilon}$.
This ensures that all the switches found in Figure~\ref{fig:epsballs1}
are registered while it avoids
false positive detections of switching, as (for instance) may occur on a
trajectories moving between $P_1^1$ and $P_2^1$ passing very
close to $P_2^9$ or $P_2^2$ (see Figure (\ref{fig_man})). One can then use simple
logic gates to connect up outputs and inputs to form a computational system.

\begin{figure}
\centerline{\mbox{\psfig{file=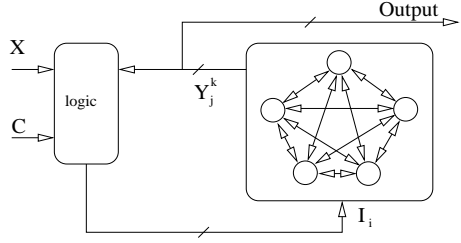,width=6cm}}}
\caption{\label{fig_counter}
Schematic wiring diagram for using the five-oscillator system for
finite-state computation. The logic
circuit steers the states around two possible periodic sequences of
$P_j^k$ depending on initial condition. The input to the system
$X$ and clock $C$ are connected to one of the $I_i$ depending only on
which of the $Y_j^k$ is lowest.
}
\end{figure}

\begin{figure}
\centerline{\mbox{\psfig{file=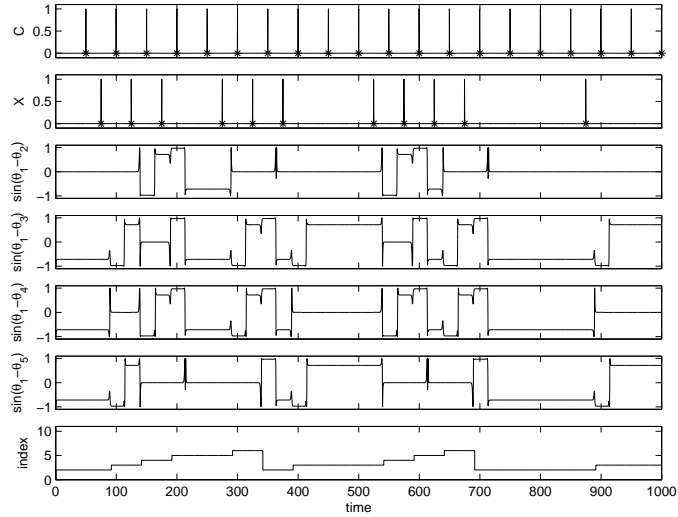,width=9cm}}}
\caption{\label{fig_simulation}
Simulation of the counter for $C(t)$ and $X(t)$ as shown in the top graphs,
for $\eta=10^{-14}$ and $\epsilon=10^{-10}$. The oscillator phase differences
$\sin(\theta_1-\theta_j)$ are shown, and the state index (bottom graph)
cycles one step through states $P_1^k$ where $k\in\{2,3,4,5,6\}$
each time an input on $X$ is received. The same output is obtained robustly
for a range of parameters.
}
\end{figure}

\begin{figure}
\centerline{\mbox{\psfig{file=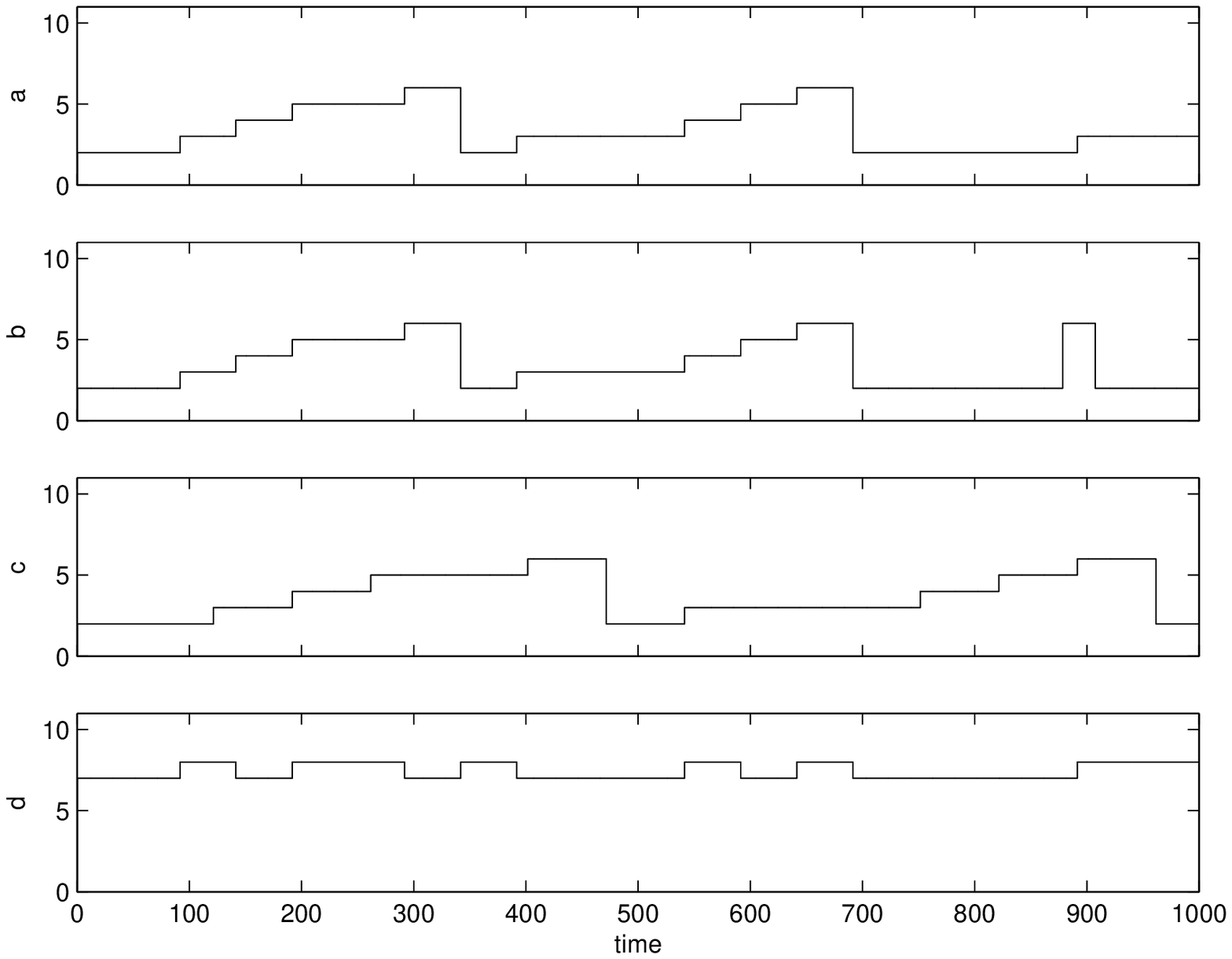,width=9cm,height=6cm}}}
\caption{\label{fig_changes}
Index giving the state of the counter for (a) the same inputs
and initial condition as Figure~\ref{fig_simulation};
(b) as for (a) but increased $\eta$
(observe the appearance of an erroneous switch just before $t=900$);
(c) as for (a) also
reliably functions with the clock rate slowed from $50$ to $70$ time units;
(d) as for (a) but starting with an initial condition near state 7; this shows
the presence of a second cycle in the network.
}
\end{figure}

As a proof-of-principal, we have used (\ref{eqhansel}) to construct a counter that
counts in either base $5$ or base $2$ depending on initial condition to the phases.
The inputs we consider are $C(t)$ a clock pulse that occurs at approximately regular
intervals and $X(t)$ an input that may or may not occur between clock pulses. We construct
the system so that pulses of $C(t)$ occurring at a state $P_2^k$ perturb the system
along the unstable manifold towards $P_1^k$ and pulses of $X(t)$ occurring at a state
$P_1^k$ perturb the system along the unstable manifold towards $P_1^{\ell(k)}$ where
$\ell=[1,3,4,5,6,2,8,7,9,10]$. In each case this corresponds to simply applying a pulse
to input $I_{p(k)}$ where $p=[4,4,1,3,2,5,3,3,2,1]$. In this way we effectively
define functions $I_i=F_1(Y_j^k)X+F_2(Y_j^k)C$
that steer the state of the system around the network rapidly and reliably even
when $\epsilon$ is very small, depending on the inputs $X$ and $C$.
We show the setup schematically in Figure~\ref{fig_counter};
Figure \ref{fig_simulation} shows the counter functioning accurately for an initial condition near
$P_1^2$. Figure~\ref{fig_changes} shows simulations on changing
parameters and in the last case, initial conditions.

There is a trade-off between on the one hand the speed of computation
and on the other hand its accuracy and sensitivity to noise. In particular the
input amplitude $\epsilon$ gives a characteristic cycle time as does the
noise amplitude $\eta$. For effective computation in the above system
we need $1\gg \epsilon\gg \eta$ and the clock cycles must lie
between the cycle times associated to the two perturbation amplitudes.

Many models use oscillators to perform computational tasks, for example
\cite{Bor&al01}. The novelty in this work is that we demonstrate how one of
the simplest possible types of coupling can give rise to robust attracting
heteroclinic networks that one can exploit to perform computations.
As in winnerless competition models \cite{Rab&al01,Afr04,Sel&al03} the average phase
differences in the globally coupled system change very little depending on
the state of the system and the computation is performed by following natural
trajectories within the system. Moreover, the state of the system is not detectable from
individual or `mean' cell outputs but is encoded globally.
By contrast, if we use asymptotically stable
attractors for storing memory states comparatively large perturbations to the
system will be needed to lift it out of the basin of attraction into another state.

The mechanism we consider is a winnerless competition model using
(a) two dimensional manifolds which allow many possibilities
to be considered at each state and so allows the
network to scale to large numbers of states without greatly increasing the
average length of path (b) the dynamics requires no careful tuning for a range of
coupling parameters, nor plasticity in the coupling.
We use two populations of processing elements (a) the globally coupled
oscillators $\theta_i$ have the function of a `memory/timing' circuit and (b)
observables $Y_j^k$ from the system that detect presence of a
particular cluster state. The latter are analogous to certain Kenyon Cells
in the mushroom body that receive inputs from an
insect antennal lobe \cite{Lau&al01,Huerta,Mar&Hug03,Per&al02} and
decode a large dense set of states in a relatively low dimensional space to a
sparse set of states in a higher dimensional space.

We believe that the simple global coupling used within the `processing element'
of our model is a positive feature.  Although symmetry is never achieved, 
the observation that the heteroclinic attractor is asymptotically stable 
means that the dynamics are well modelled even when global 
coupling is perturbed.  There may be evolutionary advantages to having 
neural subsystems that attempt to approach symmetric coupling as this can
presumably be encoded genetically very simply  (see \cite{Bre} for
other approaches to neural information processing that use
symmetries). The fact that the dynamics of the unperturbed system is 
robust means that heteroclinic attractors are model-independent and 
will exist in globally coupled systems of phase oscillators 
\cite{IzeHop,AshHets}, coupled ODE oscillators or even piecewise 
smooth systems of phase-resetting oscillators \cite{Timme}.

The dynamics we use is not present in $N\leq 4$  oscillators;
to find robust heteroclinic attractors with unstable manifolds of higher 
than one dimension we need $N\geq 5$ \cite{Ash&Bor04}. The
ideas presented should scale well on increasing $N$ beyond 5; the number of
cluster states grows exponentially but the diameter of the
heteroclinic network need not grow, meaning that rapid transition
from one state to another
is still possible.  Note we do not claim that any particular
AL/OB neural system functions via perturbations to a
robust heteroclinic attractor (though this has been suggested
\cite{Rab&al01}). We do show that a globally coupled
system of phase oscillators allows rapid and reliable computation.
 
{\bf Acknowledgments:} We acknowledge very interesting discussions
with Roman Borisyuk, David Corne, Gerhard Dangelmayr
and Marc Timme concerning this research. The research of JB is supported by
an EPSRC studentship and that of PA by a Leverhulme research fellowship. We are
very grateful for the hospitality and financial support of the MPI f\"ur
Str\"omungsforschung, G\"{o}ttingen.

\newpage

\begin{table}
$$
\begin{array}{c|c|c}
k & P_j^k & \overline{W^u(P_1^k)}\supset\\
\hline
1  & (0,0,0,\psi_j,\psi_j) & P_2^2,P_2^9,P_2^{10}\\
2  & (\psi_j,\psi_j,0,0,0) & P_2^1,P_2^3,P_2^6\\
3  & (0,0,\psi_j,0,\psi_j) & P_2^2,P_2^4,P_2^8\\
4  & (0,\psi_j,0,\psi_j,0) & P_2^3,P_2^5,P_2^{10}\\
5  & (\psi_j,0,0,0,\psi_j) & P_2^4,P_2^6,P_2^9\\
6  & (0,0,\psi_j,\psi_j,0) & P_2^2,P_2^5,P_2^7\\
7  & (0,\psi_j,0,0,\psi_j) & P_2^6,P_2^8,P_2^{10}\\
8  & (\psi_j,0,0,\psi_j,0) & P_2^3,P_2^7,P_2^9\\
9  & (0,\psi_j,\psi_j,0,0) & P_2^1,P_2^5,P_2^8\\
10 & (\psi_j,0,\psi_j,0,0) & P_2^1,P_2^4,P_2^7
\end{array}
$$
\caption{\label{tab_pos}
Representative points on the periodic orbits $P_j^k$ ($j=1,2,k=1$, $\cdots,10$)
in the unperturbed heteroclinic network (\ref{eqhansel}); parameters as in (\ref{eqn:params}).
In all  cases $\overline{W^u(P_j^k)}\supset P_{3-j}^k$, while $\overline{W^u(P_1^k)}$
contains the addition points listed ($\overline{A}$ denotes the closure of
the set $A$).}
\end{table}

\begin{thebibliography}{10}
%
\bibitem{Rab&al01}
M. Rabinovich, A. Volkovskii, P. Lecanda, R. Huerta, H. Abarbanel and G. Laurent,
\newblock {\em Phys. Rev. Lett.} {\bf 87}, 068102 (2001).

\bibitem{Bur&al02}
G.B. Burg, C.G. Galizia, R. Brandt and H. Mustaparta,
\newblock {\em J. Comp. Neurology} {\bf 446}, 123--134 (2002).

\bibitem{Mar&Hug03}
D. Martinez and E. Hugues.
\newblock {\em Proc. NATO ARW, Oct 1-2 2003, Coventry 2003} (2003).

\bibitem{Huerta}
R. Huerta, T. Nowotny, M.G. Sanchez, H.D.I. Abarbanel and M.I. Rabinovich,
\newblock {\em Neural Computation} {\bf 16}, 1601--1604 (2004).

\bibitem{Fri&al04a}
R.W. Friedrich and G. Laurent,
\newblock {\em J. Neurophysicol.} {\bf 91}, 2658--2669 (2004).

\bibitem{Fri&al04b}
R.W. Friedrich, C.J. Habermann and G. Laurent,
\newblock {\em Nature Neuroscience} {\bf 7}, 862--871 (2004).

\bibitem{Lau&al01}
G. Laurent, M. Stopfer, R.W. Friedrich, M.I. Rabinovich, A. Volkovskii and H.D.I. Abarbanel,
\newblock {\em Ann. Rev. Neurosci.} {\bf 24}, 263--297 (2001).

\bibitem{Afr04}
V. Aframovich, M. Rabinovich and P. Varona,
\newblock {\em Int. J. Bifurcation and Chaos }{\bf 14}, 4 (2004).

\bibitem{Sel&al03}
P. Seliger, L. Tsimring and M. Rabinovich,
\newblock {\em Phys. Rev. E} {\bf 67}, 011905 (2003).

\bibitem{Kru97}
M. Krupa,
\newblock {\em J. of Nonlinear Sci.} {\bf 7}, 129-176 (1997).

\bibitem{AshHets}
P. Ashwin and J.W. Swift,
\newblock {\em J. Nonlinear Sci.} {\bf 2}, 69--108 (1992);
\newblock P. Ashwin and P. Chossat,
\newblock {\em J. Nonlin. Sci.} {\bf 8}, 103--129 (1998);
\newblock P. Ashwin and M. Field,
\newblock {\em Arch. Rational Mech. Anal.} {\bf 148}, 107--143 (1999).

\bibitem{Han&al93a}
D. Hansel, G. Mato and C. Meunier,
\newblock {\em Phys. Rev. E} {\bf 48}, 3470--3477 (1993).

\bibitem{Han&al93b}
D. Hansel, G. Mato and C. Meunier,
\newblock {\em Europhys. Letts.} {\bf 23}, 367--372 (1993).

\bibitem{Kor&al01}
H. Kori and Y. Kuramoto,
\newblock {\em Phys. Rev. E} {\bf 63}, 046214 (2001).

\bibitem{Kor03}
H. Kori,
\newblock {\em Phys. Rev. E} {\bf 68}, 021919 (2003).

\bibitem{Ash&Bor04}
P. Ashwin and J. Borresen,
\newblock {\em Phys. Rev. E} {\bf 70}, 026203 (2004).

\bibitem{Armbruster}
D. Armbruster, E. Stone and V. Kirk,
\newblock {\em Chaos} {\bf 13}, 71--79 (2003).

\bibitem{Per&al02}
O. Perez-Orive, J. Mazor, G.C. Turner, S. Cassanaer, R.I. Wilson and G. Laurent,
\newblock {\em Science} {\bf 297}, 359--365 (2002).

\bibitem{Bor&al01}
R. Borisyuk, M. Denham, F. Hoppensteadt, Y. Kazanovich and O. Vinogradova,
\newblock {\em Network: Comp. Neural Syst.} {\bf 12}, 1--20 (2001).

\bibitem{Bre}
M. Breakspear,
\newblock {\em Int. J. of Neural Systems} {\bf 11}, 101--124 (2001);
\newblock D.L. Rowe,
\newblock {\em Behavioural and Brain Sciences} {\bf 24}, 827--828 (2001);
\newblock M. Breakspear, J.R. Terry and K.J. Friston.
\newblock {\em Network: computation in Neural Systems} {\bf 14}, 703--732 (2003).

\bibitem{IzeHop}
F.C. Hoppensteadt and E. Izhikevich,
\newblock {\em Weakly Connected Neural Networks}
\newblock Springer (2000).

\bibitem{Timme}
M. Timme, F. Wolf and T. Geisel,
\newblock {\em Phys. Rev. Lett.} {\bf 89}, 154105 (2002).






\end{thebibliography}
\end{document}